# Guidelines for Fast and Nondestructive Imaging in AM-AFM

Kenichi Umeda[1,2] and Noriyuki Kodera[1]

[1] *Nano Life Science Institute (WPI-NanoLSI),*

   *Kanazawa University, Kakuma-machi, Kanazawa, Ishikawa, 920-1192, Japan.*

[2] *PRESTO/JST, 4-1-8 Honcho, Kawaguchi, Saitama 332-0012, Japan.*

Corresponding Authors

Dr. Kenichi Umeda (E-mail: umeda.k@staff.kanazawa-u.ac.jp)

Prof. Noriyuki Kodera (E-mail: nkodera@staff.kanazawa-u.ac.jp)

# Abstract

Amplitude-modulation atomic force microscopy enables observation of fragile molecules at the nanometer scale. To shorten measurement times and capture dynamic molecules, increasing the frame rate is essential. Traditionally, maximum frame rates were thought to be limited by device bandwidth. However, for fragile molecules, imaging speed is often constrained by disruption from tip–sample interaction forces. Despite its significance, no comprehensive theoretical study has addressed this limitation. Here, we establish guidelines for high-speed, nondestructive AM-AFM imaging of fragile molecules. Our analysis identifies two types of forces: an impulsive force on the molecule's uphill side and a steady force linked to error saturation on the downhill side. By examining the frequency dependence of the amplitude-distance curve, we demonstrate that exciting at the resonance slope minimizes feedback error forces and allows for their easy estimation using simple equations. These findings provide valuable insights for studying fragile materials, particularly biomolecules.

# 1. Introduction

Amplitude-modulation (tapping-mode) AFM (AM-AFM) is a scanning probe technique that can measure submolecular scale surface structures and is widely used across various fields due to its compatibility with both ambient and liquid environments [1-4]. Notably, AM-AFM is highly valuable for observing fragile samples, such as biomolecules [3,5,6], because of its minimal invasiveness. In this technique, increasing the imaging rate is essential for enhancing imaging efficiency and enabling the observation of molecular dynamics. To enhance this critical capability, high-speed AFM (HS-AFM) has been developed and is now widely used for investigating biofunctional dynamics [6-15].

Conventionally, the maximum imaging rate has been assumed to be limited by the feedback bandwidth, as defined by the transfer function [7,8,16-19]. Another limitation is the signal-to-noise ratio, which degrades with an increase in feedback bandwidth [1,2,4,7,20-22]. Consequently, attempting to increase the bandwidth results in a higher minimum detectable force, which in turn increases sample damage, ultimately preventing an improvement in the maximum achievable imaging rate for fragile samples.

These limitations are based on the assumption of flat surfaces; however, additional forces arising from feedback errors during the imaging of uneven surfaces [5,8,23,24] further constrain the maximum imaging rate, particularly for fragile molecules [11,12]. Additionally, error saturation may also prevent the probe from accurately tracking the surface contours [5,8,9,17,25-29], requiring greater applied forces to obtain reliable images. Despite the critical importance of the feedback-induced forces in biomolecular imaging, a comprehensive theoretical framework has remained largely unexplored in prior research.

In formulating feedback-induced forces, it is essential to define the interaction forces at an arbitrary tip−sample distance, with the probe oscillating at a specific frequency. Although previous studies have examined this issue [3,30-32], no theoretical study has yet addressed the optimal excitation frequency for minimizing feedback-induced forces, despite the fact that force detection sensitivity varies with excitation frequency.

We have recently succeeded in quantitatively formulating the average force in AM-AFM [33], demonstrating that force detection sensitivity depends on the excitation frequency and is optimized at the resonance slope, rather than at the resonance frequency ($f_0$). However, while excitation at $f_0$ is often preferred [34-37] due to its simplicity in experimental and theoretical contexts for quantitative measurements, off-resonance excitation with peak force detection has also been used to effectively suppress dynamic forces [38]. Therefore, the frequency dependence of feedback-induced force in AM-AFM is also of significant interest.

In this study, we conduct a theoretical analysis to investigate the relationship between feedback-induced forces and the scanning velocity. For quantitative analysis, we further examine the influence of driving frequency on forces arising from feedback errors and identify that these forces are minimized at the resonance frequencies, where force sensitivity is maximized. The findings from this study provide important guidance for rapid, non-destructive observation of fragile molecules.

## 2. Two Types of Feedback-Induced Forces

To understand the tip–sample interaction forces ($F_{ts}$) that occur during AFM observations, we analyze the typical AFM results shown in Fig. 1. For the test sample, actin filaments (F-actins) were used because they are commonly employed to verify fragile imaging conditions [11,12]. F-actins were observed under standard HS-AFM experimental conditions, with a free amplitude ($A_{free}$) of 2.95 nm and excitation at a resonance slope frequency. In each panel in Fig. 1(a), the left and right images display the simultaneously acquired topographic and feedback error images in the rightward scanning direction, respectively.

First, we analyze the topographic images. Fig. 1(a)-(1) shows the image taken with an amplitude setpoint ($A_{sp}$) of 2.8 nm$_{p-0}$ and the scan velocity ($v_{scan}$) of 100 μm/s, where clear molecular structures were visible. In Fig. 1(a)-(2), $v_{scan}$ was increased to 400 μm/s while keeping the same $A_{sp}$, resulting in an increase in the frame rate from 3.2 to 12.3 fps. In the image, significant line noises appeared, making it difficult to discern clear molecular structures. Notably, the tailing effect on the downhill side of the molecule, known as the error saturation, became prominent. In Fig. 1(a)-(3), $A_{sp}$ was reduced to 2.56 nm$_{p-0}$ to obtain a clear molecular image while maintaining this high $v_{scan}$. Although the molecule became more visible, it appeared thinner and smaller compared to that shown in Fig. 1(a)-(1) due to the increased applied force. This height reduction is clearly seen in the line profile shown in Fig. 1(b). While the molecular height in Fig. 1(b)-(1) was about 8.6 nm, it progressively decreased to 8.1 nm in Fig. 1(b)-(2) and 7.8 nm in Fig. 1(b)-(3). Continued imaging under the same condition as in Fig. 1(b)-(3) led to the gradual collapse of the molecules, as seen in Fig. 1(a)-(4).

We next analyze the amplitude feedback errors, which provide insight into the force exerted on the molecule during imaging. In Fig. 1(a)-(1), although the molecules were visible in the error image, the contrast was very faint. In Fig. 1(a)-(2), the dark contrast became stronger particular in the vicinity of the filaments. The profiles shown in Fig. 1(c) indicate the decrease in the amplitude ($A_{cl}$) occurs only on the uphill side of the molecule because the feedback system cannot respond quickly enough due to the increased $v_{scan}$. In Fig. 1(a)-(3), the overall image contrast is stronger, reflecting an intensified tip–sample interaction due to the reduced $A_{sp}$.

We next analyze the average force estimated from the amplitude value using the method formulated in our previous study [33]. Since the force is most pronounced in Fig. 1(d)-(3), these data are examined first. A steady force of approximately 19 pN was observed on the flat surface surrounding the molecules, while an additional force of around 13 pN was superimposed on the molecular uphill, resulting in a total force of about 32 pN. This indicates the presence of two types of forces: $F_{steady}$, determined by $A_{sp}$ regardless of surface structure, and $F_{impulse}$, arising from feedback errors. The maximum force applied to the molecule ($F_{MaxMol}$) is obtained by summing these forces as follows:

$$\langle F_{MaxMol} \rangle = \langle F_{steady} \rangle + \langle F_{impulse} \rangle, \tag{1}$$

In AM-AFM, since the tip oscillates in the vertical direction, the brackets are used to represent the average force over one oscillation cycle of the cantilever as follows:

$$\langle F_{ts} \rangle \equiv f_{drive} \int_0^{1/f_{drive}} F_{ts}(t) dt, \tag{2}$$

where $f_{drive}$ represents the driving frequency of the cantilever. Analyzing the other profiles similarly, in Fig. 1(d)-(1), $\langle F_{steady} \rangle$ and $\langle F_{impulse} \rangle$ were estimated to be 5 pN and 3 pN, respectively. In Fig. 1(d)-(2), $\langle F_{steady} \rangle$ remained unchanged at 5 pN; however, as $v_{scan}$ increased, the feedback error grew larger, resulting in an increase of $\langle F_{impulse} \rangle$ to 13 pN and $\langle F_{MaxMol} \rangle$ to 18 pN. In Fig. 1(d)-(3), further reducing $A_{sp}$ increases the likelihood of the molecules being damaged by the tip. This analysis indicates that the maximum $v_{scan}$ is constrained by $F_{steady}$ and $F_{impulse}$, which necessitates a quantification of these forces.

## 3. Forces Resulting from Feedback Error

The $F_{ts}$ correlates with the tip interaction depth ($\Delta z_{int}$), defined as the tip–sample distance, with the zero position set at the point where $F_{ts}$ or $A_{cl}$ begins to vary. This correlation can be approximately expressed using $\alpha_{\Delta z \to F}$, which represents the conversion coefficient from $\Delta z_{int}$ to $\langle F_{ts} \rangle$, as follows:

$$\langle F_{ts} \rangle = \alpha_{\Delta z \to F} \Delta z_{int}, \tag{3}$$

where $\langle F_{ts} \rangle$ takes a positive value, whereas $\alpha_{\Delta z \to F}$ and $\Delta z_{int}$ take negative values. We will derive an analytical expression of $\alpha_{\Delta z \to F}$ later. Then, $F_{steady}$ and $F_{impulse}$ can also be decomposed into terms of $\Delta z_{int}$ as follows:

$$\begin{aligned} \langle F_{steady} \rangle &= \alpha_{\Delta z \to F} \Delta z_{steady}, \\ \langle F_{impulse} \rangle &= \alpha_{\Delta z \to F} \Delta z_{impulse}, \end{aligned} \tag{4}$$

where $\Delta z_{steady}$ and $\Delta z_{impulse}$ are defined as $\Delta z_{int}$ resulting from $F_{steady}$ and $F_{impulse}$, respectively.

When a molecule has a circular cross-section smaller than the radius of the tip curvature, its apparent shape approximately reflects the tip shape. In Fig. 2(a), by approximating with a quadratic function, the surface topography ($z_{surf}$) is represented as follows:

$$z_{surf}(x) = \begin{cases} h_{mol} - 4\dfrac{h_{mol}}{w_{mol}^2}(x - x_0)^2 & \text{on molecular region}, \\ 0 & \text{otherwise}, \end{cases} \tag{5}$$

where $h_{mol}$, $w_{mol}$, and $x_0$ represent the height, the apparent width, and the central position of the molecule, respectively.

We will now discuss the feedback error encountered during the tip scan [5,23,24,39]. Hereafter, we omit the equations for the regions outside the molecular area. In the low-frequency region, the phase response of the feedback system is nearly linear. Therefore, when the scan frequency is sufficiently lower than the feedback bandwidth, the tip trajectory ($z_{\text{traj}}$) can be approximated using a phase shift as follows:

$$z_{\text{traj}}(x) = h_{\text{mol}} - 4\frac{h_{\text{mol}}}{w_{\text{mol}}^2}(x - x_0 - \Delta x_{\text{FB}})^2, \tag{6}$$

where $\Delta x_{\text{FB}}$ represents the shift in the apparent molecular position resulting from the feedback error (Fig. 2(a)). Consequently, $\Delta z_{\text{impulse}}$ can be obtained by the difference between $z_{\text{surf}}$ and $z_{\text{traj}}$ as follows:

$$\begin{aligned}\Delta z_{\text{impulse}}(x) &= z_{\text{surf}}(x) - z_{\text{traj}}(x) \\ &= -4\frac{h_{\text{mol}}}{w_{\text{mol}}^2}\left[-2(x - x_0)\Delta x_{\text{FB}} + \Delta x_{\text{FB}}^2\right].\end{aligned} \tag{7}$$

From Fig. 2(a), the position at which the feedback error reaches its maximum ($x_{\text{impulse}}$) can be geometrically obtained as follows:

$$x_{\text{impulse}} = \arg\min\left[\Delta z_{\text{impulse}}(x)\right] = x_0 - \frac{w_{\text{mol}}}{2} + \Delta x_{\text{FB}}. \tag{8}$$

Therefore, $\Delta z_{\text{impulse}}$ can be explicitly calculated as follows (Fig. 2(b)):

$$\Delta z_{\text{impulse}}(x) = -4h_{\text{mol}}\left[\frac{\Delta x_{\text{FB}}}{w_{\text{mol}}} - \left(\frac{\Delta x_{\text{FB}}}{w_{\text{mol}}}\right)^2\right], \tag{9}$$

By substituting this equation into Eq. (4), $\langle F_{\text{impulse}}\rangle$ is obtained as follows (Fig. 2(c)):

$$\langle F_{\text{impulse}}\rangle = -\alpha_{\Delta z \to F}\, 4h_{\text{mol}}\left[\frac{\Delta x_{\text{FB}}}{w_{\text{mol}}} - \left(\frac{\Delta x_{\text{FB}}}{w_{\text{mol}}}\right)^2\right], \tag{10}$$

This equation is plotted in Fig. 2(d) using typical HS-AFM parameters for F-actin observation: $\alpha_{\Delta z \to F}$ = 0.031 N/m, $h_{\text{mol}}$ = 8.6 nm, and $w_{\text{mol}}$ = 27 nm, which reveals that $\langle F_{\text{impulse}}\rangle$ is proportional to $\Delta x_{\text{FB}}$ initially but begins to decrease when the second quadratic term exceeds the first term. However, at such large values of $\Delta x_{\text{FB}}$, the force should be saturated or extrapolated because this approximation model no longer holds. Consequently, only the first term should be considered as follows:

$$\langle F_{\text{impulse}}\rangle \approx -\alpha_{\Delta z \to F}\, 4h_{\text{mol}}\frac{\Delta x_{\text{FB}}}{w_{\text{mol}}}. \tag{11}$$

To find $\Delta x_{\text{max}}$, the value of $\Delta x_{\text{FB}}$ at which the force begins to saturate, we must solve the equation for

$\Delta x_{FB}$ below:

$$\frac{d}{d\Delta x_{FB}}\langle F_{impulse}\rangle = -\alpha_{\Delta z \to F} 4\frac{h_{mol}}{w_{mol}}\left(1-\frac{2}{w_{mol}}\Delta x_{FB}\right) = 0, \tag{12}$$

which yields the expression of $\Delta x_{max}$ as follows:

$$\Delta x_{max} = \frac{w_{mol}}{2}. \tag{13}$$

Substituting this equation into Eq. (10), the upper limit of $\langle F_{impulse}\rangle$ ($\langle F_{limit}\rangle$) is obtained as follows

$$\langle F_{limit}\rangle = -\alpha_{\Delta z \to F} h_{mol}. \tag{14}$$

Also, $\Delta x_{FB}$ is calculated from $\Delta\theta_{FB}$, which represents the phase delay of the molecular position in the scan period as follows:

$$\Delta x_{FB} = \frac{\Delta\theta_{FB}}{\pi} W_{scan}, \tag{15}$$

where $W_{scan}$ represents the horizontal scan width. Since the feedback bandwidth $B_{FB45°}$ is defined as the frequency at which the phase delay in the feedback signal with respect to the topographical variations becomes $-45°$ [11,15], $\Delta\theta_{FB}$ can be obtained as follows:

$$\Delta\theta_{FB} = \frac{\pi}{4}\frac{f_{scan}}{B_{FB45°}}, \tag{16}$$

where the scan frequency $f_{scan}$ is calculated as follows:

$$f_{scan} = \frac{v_{scan}}{2W_{scan}}. \tag{17}$$

Combining the above three equations, $\Delta x_{FB}$ can be calculated as follows:

$$\Delta x_{FB} = \frac{1}{8}\frac{v_{scan}}{B_{FB45°}}. \tag{18}$$

Substituting $\Delta x_{FB}$ into Eq. (11) yields $F_{impulse}$ as follows:

$$\langle F_{impulse}\rangle \approx -\alpha_{\Delta z \to F}\frac{h_{mol}}{2w_{mol}}\frac{v_{scan}}{B_{FB45°}}. \tag{19}$$

Using Eq. (18), we convert $\Delta x_{FB}$ on the horizontal axis of Fig. 2(d) to $v_{scan}$ at $B_{FB45°} = 60$ kHz, which is a typical HS-AFM parameter, and find that the linear approximation deviates from the original equation when $v_{\text{scan}}$ exceeds 2000 μm/s, which is an unrealistic velocity in this $B_{\text{FB45°}}$. This indicates that the linear approximation holds under most experimental conditions.

Furthermore, in HS-AFM experiments, target molecules are often weakly adsorbed onto the substrate to visualize their dynamics. In such cases, it has been empirically demonstrated that a molecule is slightly resistant to disruption caused by the scanning effect. This phenomenon may be attributed to the slight displacement of the molecule due to being pushed by the tip interaction, which effectively suppresses $F_{\text{impulse}}$. To account for this effect, the coefficient $\xi_{\text{mol}}$ is incorporated as follows:

$$\left\langle F_{\text{impulse}} \right\rangle \approx -\alpha_{\Delta z \to F} \xi_{\text{mol}} \frac{h_{\text{mol}}}{2w_{\text{mol}}} \frac{v_{\text{scan}}}{B_{\text{FB45°}}}. \tag{20}$$

The $\xi_{\text{mol}}$ may depend on $v_{\text{scan}}$ as well as the substrate interaction. In this way, we have succeeded in formulating the forces related to feedback error.

## 4. Forces Required to Eliminate Error Saturation

In this section, we will discuss the error saturation [5,7,8,17,23,24], which is prominently observed in Fig. 1(a)-(2). As illustrated in Fig. 2(e), a trajectory tailing frequently manifests at the downhill side of a molecule because the feedback error saturates at the maximum value of $A_\text{sp} - A_\text{free}$ when the tip is instantaneously displaced from the sample surface. Although it does not result in sample damage, it should be suppressed as much as possible, since information about the surface topography is significantly lost.

Therefore, we next theoretically formulate the error saturation, which also limits the maximum $v_\text{scan}$. Since the effect occurs when $F_\text{impulse}$ becomes smaller than $F_\text{steady}$ (Fig. 2(f)), $x_\text{ES}$, the position where the error saturation begins, can be determined by solving the following equation for $x$:

$$\langle F_\text{MaxMol} \rangle = \langle F_\text{steady} \rangle + \langle F_\text{impulse} \rangle = 0. \tag{21}$$

Substituting Eq. (4) into this equation yields

$$\Delta z_\text{impulse}(x) = -\frac{\langle F_\text{steady} \rangle}{\alpha_{\Delta z \to F}}. \tag{22}$$

Substituting Eq. (7) into this equation and solving for $x$ yields $x_\text{ES}$ as follows:

$$x_\text{ES} = x_0 - \frac{1}{2\mu_\text{mol}} \frac{\langle F_\text{steady} \rangle}{\alpha_{\Delta z \to F} \Delta x} + \frac{\Delta x_\text{FB}}{2},$$
$$\text{where} \quad \mu_\text{mol} = \frac{4 h_\text{mol}}{w_\text{mol}^2}. \tag{23}$$

Substituting this equation into Eq. (6) yields $h_\text{ES}$, the height where the error saturation begins (Fig. 2(e)), as follows:

$$h_\text{ES} \equiv z_\text{traj}(x = x_\text{ES})$$
$$= h_\text{mol} - \mu_\text{mol} \left( -\frac{1}{2\mu_\text{mol}} \frac{\langle F_\text{steady} \rangle}{\alpha_{\Delta z \to F} \Delta x_\text{FB}} - \frac{\Delta x_\text{FB}}{2} \right)^2. \tag{24}$$

The error saturation can be simulated by linearly extrapolating a tip trajectory at $x = x_{ES}$. For this, we must calculate the derivative of the trajectory as follows:

$$z'_{traj}(x) = \frac{dz_{traj}(x)}{dx} = -2\mu(x - x_0 - \Delta x_{FB}). \tag{25}$$

Consequently, the trajectory gradient at $x = x_{ES}$ is calculated as

$$z'_{ES} = z'_{tip}(x_{ES}) = \frac{\langle F_{steady} \rangle}{\alpha_{\Delta z \to F} \Delta x_{FB}} + \mu \Delta x_{FB}. \tag{26}$$

Using this equation, the tip trajectory accounting for the error saturation ($z_{traj,ES}$) is calculated as follows:

$$z_{traj,ES}(x) = \begin{cases} h_{mol} - \mu_{mol}(x - x_0 - \Delta x_{FB})^2 & \text{if } x < x_{ES}, \\ z'_{ES}(x - x_{ES}) + h_{ES} & \text{otherwise.} \end{cases} \tag{27}$$

Since $h_{ES}$ takes values in the range of 0 to $h_{mol}$ and is proportional to $\langle F_{steady} \rangle^2$, as seen in Eq. (24), we found that the amount of the error saturation can be more easily adjusted using a characteristic variable $\chi_{ES}$, defined as follows:

$$\chi_{ES} \equiv \sqrt{1 - \frac{h_{ES}}{h_{mol}}} \quad (0 \leq \chi_{ES} \leq 1),$$
$$\therefore h_{ES} = h_{mol}(1 - \chi_{ES}^2). \tag{28}$$

Substituting this equation in to Eq. (27) yields $z_{traj,ES}$ as follows:

$$z_{traj,ES}(x) = \begin{cases} h_{mol} - \mu_{mol}(x - x_0 - \Delta x_{FB})^2 & \text{if } x < x_{ES}, \\ z'_{ES}(x - x_{ES}) + h_{mol}(1 - \chi_{ES}^2) & \text{otherwise.} \end{cases} \tag{29}$$

This equation is plotted at various $\chi_{ES}$ values in Fig. 2(g) by assuming the observation of F-actin [11]: $h_{mol}$ = 8.6 nm, $w_{mol}$ = 27 nm, $B_{FB45°}$ = 20 kHz, $v_{scan}$ = 100 μm/s, $W_{scan}$ = 200 nm, and $\langle F_{steady} \rangle$ = 10 pN. As shown, we confirmed that the width of the error saturation region narrows as $\chi_{ES}$ increases.

Furthermore, solving Eq. (23) for $\langle F_{steady} \rangle$ yields

$$\langle F_{\text{steady}} \rangle = -\alpha_{\Delta z \to F} \Delta x_{\text{FB}} \left( \mu_{\text{mol}} \Delta x_{\text{FB}} + 2\sqrt{h_{\text{mol}} \mu_{\text{mol}} \left(1 - \frac{h_{\text{ES}}}{h_{\text{mol}}}\right)} \right)$$
$$= -\alpha_{\Delta z \to F} 4 h_{\text{mol}} \left[ \sqrt{1 - \frac{h_{\text{ES}}}{h_{\text{mol}}}} \left(\frac{\Delta x_{\text{FB}}}{w_{\text{mol}}}\right) + \left(\frac{\Delta x_{\text{FB}}}{w_{\text{mol}}}\right)^2 \right]. \qquad (30)$$

Substituting Eq. (28) into this equation yields a criterion equation as follows:

$$\langle F_{\text{steady}} \rangle = -\alpha_{\Delta z \to F} 4 h_{\text{mol}} \left[ \chi_{\text{ES}} \left(\frac{\Delta x_{\text{FB}}}{w_{\text{mol}}}\right) + \left(\frac{\Delta x_{\text{FB}}}{w_{\text{mol}}}\right)^2 \right]. \qquad (31)$$

Similar to the previous section, by neglecting the second term and substituting Eq. (18), we obtain the equation below:

$$\langle F_{\text{steady}} \rangle \approx -\alpha_{\Delta z \to F} \chi_{\text{ES}} \frac{h_{\text{mol}}}{2 w_{\text{mol}}} \frac{v_{\text{scan}}}{B_{\text{FB45°}}}. \qquad (32)$$

This equation means that suppressing the error saturation requires an increase in $F_{\text{steady}}$, which consequently leads to greater sample damage. Therefore, as demonstrated in Fig. 1, imaging of fragile biomolecules is generally conducted while allowing a certain degree of error saturation. Alternatively, several techniques have been proposed to suppress error saturation without increasing $F_{\text{steady}}$ [5,8,9,25-29].

By combining Eqs. (20) and (32), $F_{\text{MaxMol}}$ can be expressed as follows:

$$\langle F_{\text{MaxMol}} \rangle = -\left(\xi_{\text{mol}} + \chi_{\text{ES}}\right) \alpha_{\Delta z \to F} \frac{h_{\text{mol}}}{2 w_{\text{mol}}} \frac{v_{\text{scan}}}{B_{\text{FB45°}}}. \qquad (33)$$

This equation indicates that $\chi_{\text{ES}}$ determines the ratio of $F_{\text{steady}}$ to $F_{\text{impulse}}$. From Fig. 2(c), typically, an allowable range is $\chi_{\text{ES}} = 0.5–0.7$, which means that $F_{\text{steady}}$ is smaller than $F_{\text{impulse}}$. Similar to $\langle F_{\text{impulse}} \rangle$, Eq. (33) should be saturated by $\langle F_{\text{limit}} \rangle$ in Eq. (14).

Conversely, we can estimate the maximum $v_{\text{scan}}$ by substituting the previously known biomolecular binding force into $\langle F_{\text{MaxMol}} \rangle$ as follows:

$$v_{\text{scan}} = -\frac{B_{\text{FB45}°}}{(\xi_{\text{mol}} + \chi_{\text{ES}})\alpha_{\Delta z \to F}} \frac{2w_{\text{mol}}}{h_{\text{mol}}} \langle F_{\text{MaxMol}} \rangle. \tag{34}$$

Note that molecular breakage is a stochastic phenomenon; therefore, the probability increases with imaging time, even at constant force.

## 5. Conversion of Interaction Depth to Amplitude

To apply the derived equations to practical experiments, we derive below an analytical solution for $\alpha_{\Delta z \to F}$ based on the characteristic that $\Delta z_{int}$, $\Delta A_{ts}$, and $\langle F_{ts} \rangle$ are correlated with each other (Fig. 3(a)). As we previously revealed [33], when exciting the cantilever at the resonant slope, $\langle F_{ts} \rangle$ and $\Delta A_{ts}$ can be related by a linear approximation as follows:

$$\langle F_{ts} \rangle = \alpha_{\Delta A \to F} \Delta A_{ts} \tag{35}$$

where $\alpha_{\Delta A \to F}$ represents the conversion coefficient from $\Delta A_{ts}$ to $F_{ts}$. Moreover, $\Delta A_{ts}$ can also be obtained from $\Delta z_{int}$ through a linear approximation as follows:

$$\Delta A_{ts} = \alpha_{\Delta z \to \Delta A} \Delta z_{int}, \tag{36}$$

where $\alpha_{\Delta z \to \Delta A}$ represents the conversion coefficient from $\Delta z_{int}$ to $\Delta A_{ts}$, which ranges from 0.5 to 1. As the tip−sample distance decreases, only the peak bottom ($\Delta A_{bot}$) decreases for $\alpha_{\Delta z \to \Delta A} = 0.5$, whereas both the peak top ($\Delta A_{top}$) and $\Delta A_{bot}$ decrease for $\alpha_{\Delta z \to \Delta A} = 1$. Accordingly, $\langle F_{ts} \rangle$ can be calculated from $\Delta z_{int}$ as follows:

$$\begin{aligned}\langle F_{ts} \rangle &= \left( \alpha_{\Delta z \to \Delta A} \cdot \alpha_{\Delta A \to F} \right) \Delta z_{int} \\ &= \alpha_{\Delta z \to F} \Delta z_{int},\end{aligned} \tag{37}$$

Based on this equation, we derive $\alpha_{\Delta z \to \Delta A}$ in order to obtain $\alpha_{\Delta z \to F}$ below.

Although AM-AFM can operate in both repulsive and attractive regimes, we focus our analysis on the repulsive regimes ($\langle F_{ts} \rangle > 0$), because molecular disruptions are mainly caused by repulsive forces. As illustrated in Fig. 3(b), $\Delta A_{ts}$ can be calculated by the average of $\Delta A_{top}$ and $\Delta A_{bot}$ as follows;

$$\begin{aligned}\Delta A_{ts} &= \frac{1}{2}\left( \Delta A_{bot} + \Delta A_{top} \right) \\ \therefore \Delta A_{top} &= 2\Delta A_{ts} - \Delta A_{bot}.\end{aligned} \tag{38}$$

Meanwhile, $\Delta\langle z_{\text{tip}}\rangle$, the deviation of the average cantilever displacement resulting from $F_{\text{ts}}$, can be calculated from the difference between $\Delta A_{\text{top}}$ and $\Delta A_{\text{bot}}$ as follows:

$$\Delta\langle z_{\text{tip}}\rangle = \frac{1}{2}\left[A_{\text{free}} + \Delta A_{\text{top}} - \left(A_{\text{free}} + \Delta A_{\text{bot}}\right)\right] \qquad (39)$$
$$= \frac{1}{2}\left(\Delta A_{\text{top}} - \Delta A_{\text{bot}}\right).$$

Note that $\Delta A_{\text{ts}}$ and $\Delta\langle z_{\text{tip}}\rangle$ take negative and positive values, respectively. Combining these equations, we obtain

$$\Delta\langle z_{\text{tip}}\rangle = \Delta A_{\text{ts}} - \Delta A_{\text{bot}}. \qquad (40)$$

Simulation analysis revealed that when the Young's modulus of the sample is several hundred MPa or more, $\Delta A_{\text{bot}}$ approximately equals $\Delta z_{\text{int}}$ as follows:

$$\Delta A_{\text{bot}} \approx \Delta z_{\text{int}}. \qquad (41)$$

This approximation holds true for any $Q_{\text{cl}}$ and $f_{\text{drive}}$. Therefore, we obtain

$$\Delta A_{\text{ts}} = \Delta z_{\text{int}} + \Delta\langle z_{\text{tip}}\rangle. \qquad (42)$$

Meanwhile, $\Delta\langle z_{\text{tip}}\rangle$ can be calculated from the average force divided by $k_{\text{cl}}$, the spring constant of the cantilever, as follows:

$$\Delta\langle z_{\text{tip}}\rangle = \frac{\langle F_{\text{ts}}\rangle}{k_{\text{cl}}} = \frac{\alpha_{\Delta A \to F}}{k_{\text{cl}}}\Delta A_{\text{ts}}, \qquad (43)$$

Combining Eqs. (42) and (43), we obtain the equation below:

$$\Delta A_{\text{ts}} = \left(1 - \frac{\alpha_{\Delta A \to F}}{k_{\text{cl}}}\right)^{-1}\Delta z_{\text{int}} = \alpha_{\Delta z \to \Delta A}\Delta z_{\text{int}}. \qquad (44)$$

The analytical solution of $\alpha_{\Delta A \to F}$ was derived in previous research as follows [33]:

$$\alpha_{\Delta A \to F} = -\frac{k_{\text{cl}}}{2Q_{\text{cl}}}\left[\frac{\tilde{\omega}_{\text{drive}}^2}{Q_{\text{cl}}\left(1 - \tilde{\omega}_{\text{drive}}^2\right)} + Q_{\text{cl}}\left(1 - \tilde{\omega}_{\text{drive}}^2\right)\right] \quad (\tilde{\omega}_{\text{drive}} < 1), \qquad (45)$$

where $\tilde{\omega}_{\text{drive}}$ is a normalized $f_{\text{drive}}$, defined as follows:

$$\tilde{\omega}_{\text{drive}} \equiv \frac{\omega_{\text{drive}}}{\omega_0} = \frac{f_{\text{drive}}}{f_0}. \tag{46}$$

In Fig. 3(c), as we previously reported [33], the frequency dependence of $\alpha_{\Delta A \to F}$ shows a minimum at a resonance slope frequency, which is referred to as the MinForce frequency. By exciting the cantilever at this frequency, a maximum change in amplitude can be achieved even with a minimal change in force. Although two MinForces are present on both the lower and upper sides of the resonance peak, only the lower MinForce frequency ($f_{\text{LMF}}$) must be considered for repulsive forces, which can be expressed as follows:

$$\tilde{\omega}_{\text{LMF}} \equiv \frac{f_{\text{LMF}}}{f_0} = \sqrt{1 - \frac{1}{Q_{\text{cl}}}} \quad (Q_{\text{cl}} > 1). \tag{47}$$

By substituting this equation to Eq. (45), we obtain a simplified analytical equation as follows:

$$\alpha_{\Delta A \to F}\big|_{\text{LMF}} = -\frac{k_z}{2Q_{\text{cl}}}\left(2 - \frac{1}{Q_{\text{cl}}}\right) \quad (Q_{\text{cl}} > 1). \tag{48}$$

In Fig. 3(d), $\alpha_{\Delta A \to F}$ decreases inversely with $Q_{\text{cl}}$, indicating that higher sensitivity can be obtained with increasing $Q_{\text{cl}}$.

As previously investigated [1,33], conventionally, it has been postulated that the maximum force sensitivity can be achieved by exciting at the frequency where the amplitude slope is at its maximum, which is referred to as MaxSlope. In Fig. 3(d), utilizing the approximate equation derived in the previous study [33], the result of MaxSlope is also plotted, which is nearly equivalent to MinForce.

Substituting Eq. (45) into Eq. (44) yields

$$\alpha_{\Delta z \to \Delta A} = \left\{ 1 + \frac{1}{2Q_{\text{cl}}}\left[ \frac{\tilde{\omega}_{\text{drive}}^2}{Q_{\text{cl}}\left(1 - \tilde{\omega}_{\text{drive}}^2\right)} + Q_{\text{cl}}\left(1 - \tilde{\omega}_{\text{drive}}^2\right) \right] \right\}^{-1} \quad (\tilde{\omega}_{\text{drive}} < 1). \tag{49}$$

In Fig. 3(e), the $\tilde{\omega}_{\text{drive}}$ dependence shows a maximum at the MinForce frequency, rather than the minimum observed in $\alpha_{\Delta A \to F}$. By substituting Eq. (47) into this equation, the analytical solution at MinForce is obtained as follows:

$$\alpha_{\Delta z \to \Delta A}\big|_{\text{LMF}} = \frac{1}{1 + \dfrac{1}{2Q_{\text{cl}}}\left(2 - \dfrac{1}{Q_{\text{cl}}}\right)} \quad (Q_{\text{cl}} > 1). \tag{50}$$

In Fig. 3(f), $\alpha_{\Delta z \to \Delta A}$ varies between 0.65 and 1, depending on $Q_{\text{cl}}$. As $Q_{\text{cl}}$ decreases, $\alpha_{\Delta z \to \Delta A}$ approaches 0.65, indicating that $\Delta A_{\text{top}}$ is only slightly reduced by the substrate interaction. In contrast, at high $Q_{\text{cl}}$, $\alpha_{\Delta z \to \Delta A}$ asymptotically approaches 1, as both $\Delta A_{\text{bot}}$ and $\Delta A_{\text{top}}$ decrease simultaneously. The MaxSlope result is nearly identical to that of MinForce.

Furthermore, multiplying Eq. (45) and Eq. (49) yields the analytical expression for $\alpha_{\Delta z \to F}$ as follows:

$$\begin{aligned}\alpha_{\Delta z \to F} &= \alpha_{\Delta z \to \Delta A} \cdot \alpha_{\Delta A \to F} \\ &= \frac{-k_{\text{cl}}}{1 + 2Q_{\text{cl}}\left[\dfrac{\tilde{\omega}_{\text{drive}}^2}{Q_{\text{cl}}(1-\tilde{\omega}_{\text{drive}}^2)} + Q_{\text{cl}}(1-\tilde{\omega}_{\text{drive}}^2)\right]^{-1}} \quad (\tilde{\omega}_{\text{drive}} \leq 1).\end{aligned} \tag{51}$$

Substituting Eq. (47) into this equation gives the analytical solution at MinForce as follows:

$$\alpha_{\Delta z \to F}\big|_{\text{LMF}} = \frac{-k_{\text{cl}}}{1 + 2Q_{\text{cl}}\left(2 - \dfrac{1}{Q_{\text{cl}}}\right)^{-1}} \quad (Q_{\text{cl}} > 1). \tag{52}$$

In static-mode AFM, $\alpha_{\Delta z \to F}$ corresponds to $k_{\text{cl}}$. In Fig. 3(g,h), both the frequency and $Q_{\text{cl}}$ dependences exhibit a trend similar to that observed for $\alpha_{\Delta A \to F}$. The results show that exciting at the MinForce frequency reduces $\alpha_{\Delta A \to F}$ to approximately $0.31 \cdot k_{\text{cl}}$, $0.15 \cdot k_{\text{cl}}$, and $0.046 \cdot k_{\text{cl}}$ for $Q_{\text{cl}}$ = 1.5, 5, and 20, respectively. Under these conditions, feedback-error-induced sample damage can be minimized. Notably, at $\tilde{\omega}_{\text{drive}} = 1$, $\alpha_{\Delta z \to F}$ converges to $k_{\text{cl}}$, where the feedback-error force matches the magnitude observed in static-mode AFM.

Particularly, assuming a typical HS-AFM parameter of $Q_{\text{cl}}$ = 1.5 and excitation at the MinForce frequency, $\langle F_{\text{ts}} \rangle$ can be readily estimated as follows:

$$\langle F_{\text{ts}} \rangle = -0.308 \cdot k_{\text{cl}} \Delta z_{\text{int}}. \tag{53}$$

This equation indicates that feedback-induced force can be reduced to 1/3 of the static-mode AFM by exciting the cantilever at the MinForce frequency.

## 6. Numerical Simulation-based Validation of Theory

To validate the derived equations, we conducted numerical simulations based on the Hertzian model as follows [30,33]:

$$F_{\text{Hertz}}(\delta) = \frac{4E^*}{3}\sqrt{R_{\text{tip}}}\left(-\delta\right)^{3/2}, \tag{54}$$

where $R_{\text{tip}}$, $\delta$, and $E^*$ are the tip's radius of curvature, indentation depth, and reduced Young's modulus, respectively. When the tip's Young's modulus is sufficiently higher than that of the sample, $E^*$ approximates the sample's Young's modulus [33]. The simulation conditions used typical HS-AFM parameters: $k_{\text{cl}} = 0.1$ N/m and $A_{\text{free}} = 0.3$ nm, with other parameters identical to those in the previous study [33]. In Fig. 4(a,b), the analytical result (yellow broken line) generally agrees well with the simulation results for both force and amplitude curves when $E^*$ is 100 MPa or higher. This range includes the $E^*$ values of most structural proteins, typically on the order of GPa. However, when $E^*$ is 10 MPa or lower, the approximation overestimates the force.

We also compare the results when excited at $f_0$ in Fig. 4(c,d). When $E^*$ exceeds 100 MPa, the force increases more steeply than the analytical MinForce result (yellow broken line) at a shallow $\Delta z_{\text{int}}$ of ~1 nm, after which the curve exhibits a shoulder-like profile (Fig. 4(c)). As described in Fig. 3(f), the analytical calculations indicate that $\alpha_{\Delta z \to F}$ matches $k_{\text{cl}}$ when excited at $f_0$. To verify this, a line with a gradient of $k_{\text{cl}}$ is superimposed from $\Delta z_{\text{int}} = 0$ (right cyan broken line), confirming that the force rise at $\Delta z_{\text{int}} = 0$ aligns with the gradient of $k_{\text{cl}}$. Meanwhile, the amplitude decreases almost linearly as $\Delta z_{\text{int}}$ decreases (Fig. 4(d)), similarly to the analytical MinForce solution. The negligible difference between the amplitude curves at MinForce and $f_0$ suggests that the force cannot be accurately determined solely based on the amplitude curves, necessitating the prior recording of $f_{\text{drive}}$.

Furthermore, in Fig. 4(a,c), when $\Delta z_{\text{int}}$ is reduced to less than $-A_{\text{free}}$, the force increases steeply and deviates significantly from the analytical result (yellow broken line). We found that the force gradient at the distance of $\Delta z_{\text{int}} \leq -A_{\text{free}}$ can be well fitted by $k_{\text{cl}}$ as follows (cyan broken line):

$$\langle F_{\text{ts}} \rangle = -k_{\text{cl}}\left(\Delta z_{\text{int}} + A_{\text{free}}\right) \quad \text{if } \Delta z_{\text{int}} \ll -A_{\text{free}}, \tag{55}$$

Therefore, the force becomes as strong as that in static-mode AFM when $\Delta z_{\text{int}} < -A_{\text{free}}$.

This analytical result is consistent with an empirical guideline of AM-AFM, i.e., to fully utilize the benefits of the dynamic mode, the cantilever must be excited to an extent comparable to the size of the molecule being observed. However, we must be cautious when further increasing the amplitude, as this is likely to increase peak forces and make $A_{\text{free}}$ susceptible to fluctuations in cantilever excitation efficiency, which necessitates a further increase in $F_{\text{steady}}$.

# 7. Experimental Validation of Theory

To ascertain the applicability of the derived theories to practical experiments, we performed force curve experiments on mica using an HS-AFM setup. Detailed experimental conditions are described in a previous study [33]. Briefly, we used a small cantilever (BLAC10DS-A2, Olympus) with $f_0$ of 682 kHz, $Q_{cl}$ of 1.8, and $k_{cl}$ of 0.13 N/m. The cantilever was dynamically excited using the piezoacoustic method with an $A_{free}$ of 2.8 ± 0.1 nm at various $f_{drive}$ ranging from 217 to 960 kHz, which corresponding to $\tilde{\omega}_{drive}$ from 0.32 to 1.41. To reduce the instrumental noise, 300 successively measured curves were integrated.

In Fig. 5(a), the experimental results showed that below $\tilde{\omega}_{peak} = 0.92$, a gradual increase in the force was observed at $\Delta z_{int} = 0$, and when $\Delta z_{int}$ fell below $A_{free}$, the distance dependency of the force became nearly linear. Above $\tilde{\omega}_{peak}$, the force curve exhibited a shoulder-like profile, as predicted by the simulation results of Fig. 4(c). In contrast, in Fig. 5(b), the distance dependence of the amplitude exhibited a marked variation as a function of $\tilde{\omega}_{drive}$. At low $\tilde{\omega}_{drive}$, the amplitude exhibited a gentle decrease, while at higher $\tilde{\omega}_{drive}$, a sharp decline is observed. Additionally, the amplitude reduction started at a greater distance from the sample surface as $\tilde{\omega}_{drive}$ increased, indicating that higher $\tilde{\omega}_{drive}$ enhances the force detection sensitivity, enabling the observation of weaker forces. However, upon exceeding $\tilde{\omega}_{peak}$, the amplitude curve exhibited a hump-shaped profile, with an initial increase followed by a decline as the distance decreased, and the onset of the amplitude reduction shifted closer to the sample surface. This means that increasing $\tilde{\omega}_{drive}$ beyond $\tilde{\omega}_{peak}$ deteriorates force detection sensitivity.

To validate quantitative agreement between the experiments and theory, we performed simulations under the same conditions as the experiments. The value of $E^*$ was set to 800 MPa, which was experimentally determined using a static-mode AFM [33]. As shown in Fig. 5(c,d), the experimental and the simulation results were in good agreement, especially when $\Delta z_{int}$ was less than –2 nm. In the experiment, the force increased gradually near $\Delta z_{int}$ = 0, whereas in the simulation, a shoulder-like increase is observed at $\tilde{\omega}_{drive} > 0.65$. Therefore, even when excited at the resonance slope, the theoretical calculation results overestimated the force gradient compared to the experiment. This discrepancy is presumably due to the dissipation effects and non-contact forces such as DLVO and hydration forces, which were not accounted for in the simulation.

Furthermore, we analyze the analytical solution. In Fig. 5(e), comparing the force gradient at $\Delta z_{int}$ = 0 revealed that the overall estimate was approximately twice as large as the experimental results. Additionally, Fig. 5(f) shows that, when comparing the amplitude at $\Delta z_{int}$ = 0, the analytical calculations overestimated the gradients in all $\tilde{\omega}_{drive}$ values. However, since it is crucial to account for the force generated due to feedback error when the tip–sample interaction occurs ($F_{steady} > 0$), the gradient should be compared at $\Delta z_{int} < 0$ nm.

Consequently, the force gradients at various setpoint ratios (SPRs) versus the excitation frequency are compared in Fig. 6. The gradient was calculated as the difference in force between the position where $A_{cl}$ reached each $A_{SP}$ and a point 0.8 nm closer to the surface (30% of $A_{free}$). For the analytical calculation, a gradient minimum occurs at the resonance slope. In experiments, for SPR = 0.9, a minimum was observed at the resonance slope; however, the data points were consistently about 60% lower than those from the analytical calculation. In contrast, for SPR = 0.8, the values showed better agreement, especially near the resonance slope. When SPR was reduced to 0.7, the data points consistently exceeded those of the analytical calculations.

In other words, although the gradient slightly varies with SPR, when excitation is performed at the resonant slope, the force can be estimated using the analytical equation with an accuracy within 60%. Given that molecular destruction is a stochastic phenomenon, this level of accuracy is considered acceptable for approximating the force. Furthermore, if the discrepancy arises from dissipation effects, $F_{ts}$ could be mitigated by increasing the cantilever's resonance frequency, which warrants further investigation in future studies.

## 8. Estimation of Maximum Scan Velocity

In accordance with the derived theories, the maximum allowable values of $v_{scan}$ and frame rate ($f_{fps}$) can be calculated based on the molecular disruption force and $B_{FB45°}$. For example, we have previously succeeded in observing myosin-V processively moving along F-actin upon the ATP hydrolysis [14]. In this observation, the imaging was performed with scan dimensions of 130 × 65 nm$^2$ (80 × 40 pixels$^2$) and frame time of 146.7 ms, which corresponds to $f_{fps}$ of 6.8 fps. Then, the $v_{scan}$ is calculated to be 70 μm/s using the equation as follows:

$$v_{scan} = 2W_{scan}N_y f_{fps}, \qquad (56)$$

where $N_y$ is the pixel number in the Y direction.

Based on these parameters, $\langle F_{MaxMol} \rangle$ is calculated as a function of $v_{scan}$ using Eq. (33) as shown in Fig. 7. We consider typical experimental parameters: $k_{cl}$ = 0.1 N/m and $Q_{cl}$ = 1.5, along with molecular dimensions for myosin-V ($h_{mol}$ = 5 nm and $w_{mol}$ = 15 nm), assuming $\xi_{mol} + \chi_{ES}$ =1.7. The force exhibits a linear increase in proportion to $v_{scan}$, indicating that scanning at a higher $v_{scan}$ inevitably exerts a greater force on the molecules. As $B_{FB45°}$ increases, the line shifts to higher $v_{scan}$, indicating that improvements in bandwidth can facilitate an increase in frame rate. In this experiment, we used a sample-and-hold type amplitude detector [11,13], estimating the $B_{FB45°}$ of this system to be approximately 60 kHz [11]. The binding force between myosin and F-actin was estimated at approximately 15 pN.

From the intersection of the lines at these values (Fig. 7), we determine that the maximum $v_{scan}$ is 100 μm/s, which is slightly higher than the value of 70 μm/s used in the experiment. Since molecular breakage is a stochastic phenomenon, leaving a small margin above the actual bonding strength makes this value reasonable. Currently, the fastest speed of our HS-AFM system is ~100 kHz, and we anticipate that it will be accelerated to approximately 750 kHz in the future [10]. Consequently, it will be feasible to scan at a $v_{scan}$ of more than 1000 μm/s.

Finally, we must remark that a molecular disruption by the tip may be a stochastic event, depending on the substrate interactions and the tip condition, which complicates quantitative estimation. Nevertheless, the equations proposed in this paper provide important guidelines for estimating the maximum scan rate.

## 9. Conclusions

In this study, we successfully formulated theoretical equations for the steady and impulsive forces resulting from feedback and scanning processing, respectively, which represent universal principles in AFM. The theory elucidated that these two forces share an identical analytical expression. Furthermore, we derived a quantitative analytical solution for the correlation between interaction depth and the tip–sample interaction force for AM-AFM, with the results that align with the simulation and experimental results, particularly when exciting at the resonance slope. By combining these equations, we proposed a method for predicting the maximum scan velocity and frame rate that can be applied to nondestructive imaging. The knowledge gained from this research facilitates faster and less invasive measurements, providing significant benefits for practical experiments.

# Figures

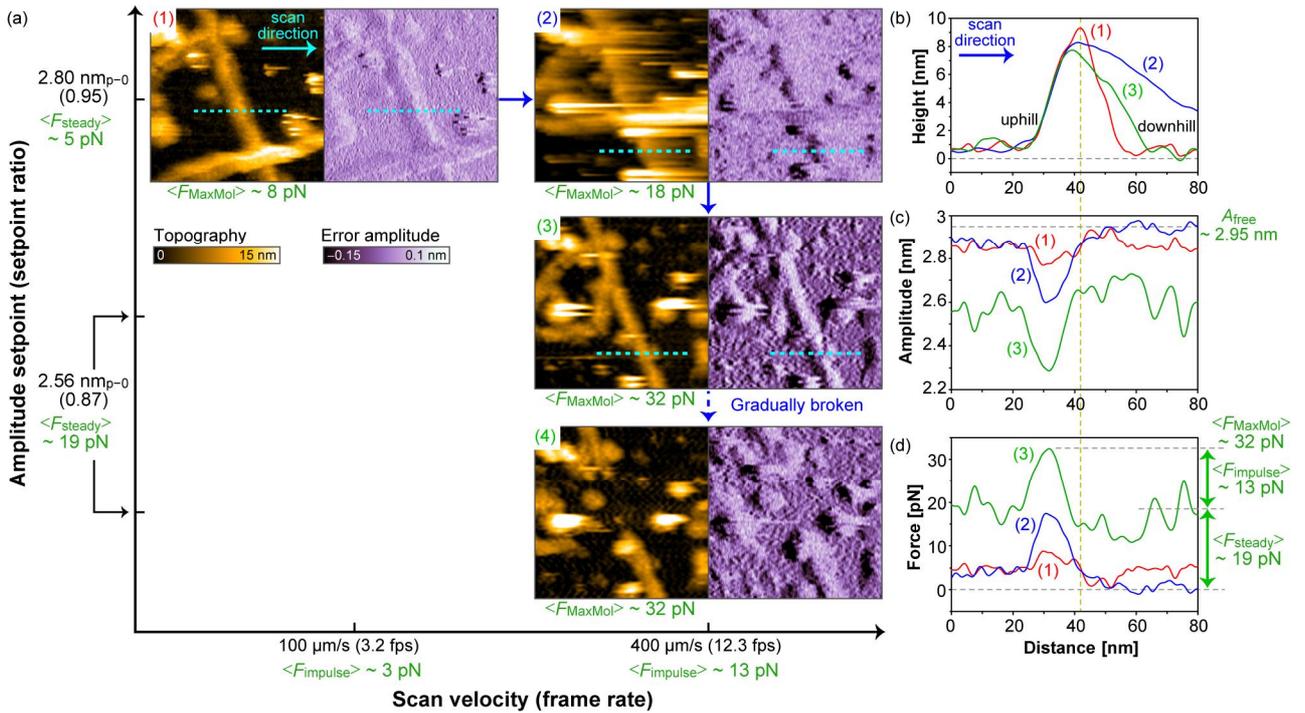

**FIG. 1.** (**a**) HS-AFM observation of F-actins that represents the relationship between $F_{steady}$, $F_{impulse}$, and $F_{MaxMol}$ as a function of amplitude setpoint and scan velocity. In each panel, the left and right images correspond to the topography and feedback error amplitude, respectively. (**b–d**) Line profiles of height (b), amplitude (c), estimated force (d), extracted from the positions indicated by the broken lines in Fig. 1(a). The amplitude signal is calculated by summing the setpoint and error amplitudes. All the images represent rightward scan data. The scans were conducted under typical HS-AFM conditions, similar to those described in Section 8, with a scan size of 150 × 150 nm$^2$.

28 / 41

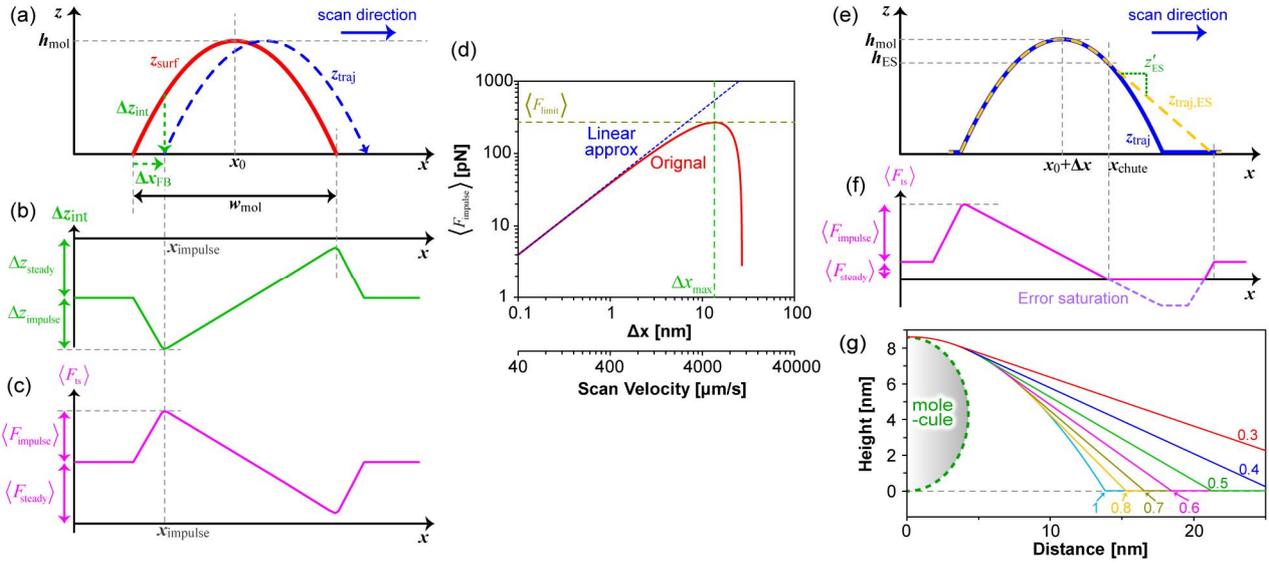

**FIG. 2.** (**a**) Molecular surface shape approximated by a quadratic function (red line, Eq. (5)) and tip trajectory not accounting for the error saturation (blue broken line, Eq. (6)); (**b**) interaction depth resulting from the feedback error (green line, Eq. (7)); (**c**) force variation resulting from the feedback error (purple line). (**d**) Relationship between $\langle F_{impulse} \rangle$ and two parameters, $\Delta x$ and scan velocity, shown with original equation (Eq. (10)), linear approximation (Eq. (11)), $\Delta x_{max}$ (Eq. (13)), and $\langle F_{limit} \rangle$ (Eq. (14)). (**e**) Tip trajectory without (Eq. (6)) and with (Eq. (27)) the error saturation; (**f**) force variation resulting from the feedback error accounting for the error saturation (purple line). (**g**) Tip trajectory accounting for error saturation dependence on $\chi_{ES}$ as the normalization variable (Eq. (29)).

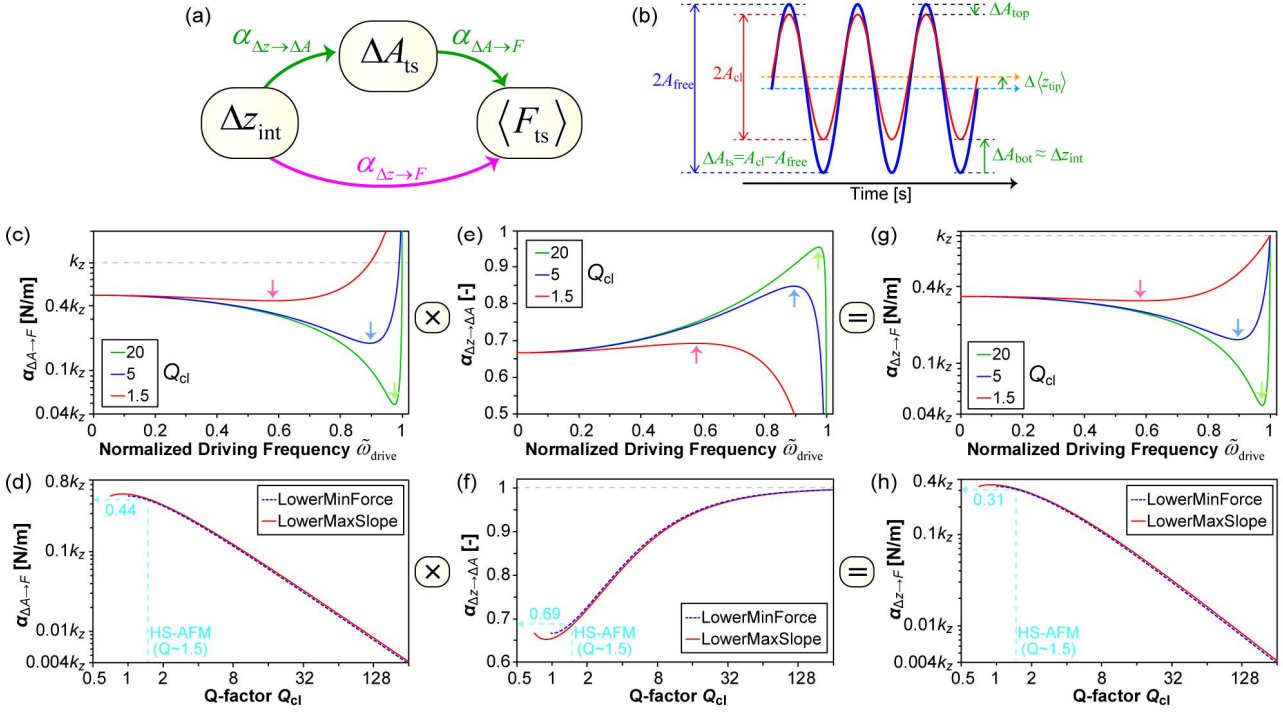

**FIG. 3.** (**a**) Schematic relationship between $\Delta z_{\mathrm{int}}$, $\Delta A_{\mathrm{ts}}$, and $\langle F_{\mathrm{ts}} \rangle$. (**b**) Schematic diagram for deriving a theoretical equation for the decrease in oscillation amplitude dependent on the interaction depth. (**c,e,g**) Driving frequency dependence of $\alpha_{\Delta A \to F}$ (c), $\alpha_{\Delta z \to \Delta A}$ (e), and $\alpha_{\Delta z \to F}$ (g) for different $Q_{\mathrm{cl}}$ values. The arrows indicate the minimum or maximum points, which corresponds to the MinForce frequencies. (**d,f,h**) $Q_{\mathrm{cl}}$ dependence of $\alpha_{\Delta A \to F}$ (d), $\alpha_{\Delta z \to \Delta A}$ (f), and $\alpha_{\Delta z \to F}$ (h) at the lower MinForce and MaxSlope frequencies.

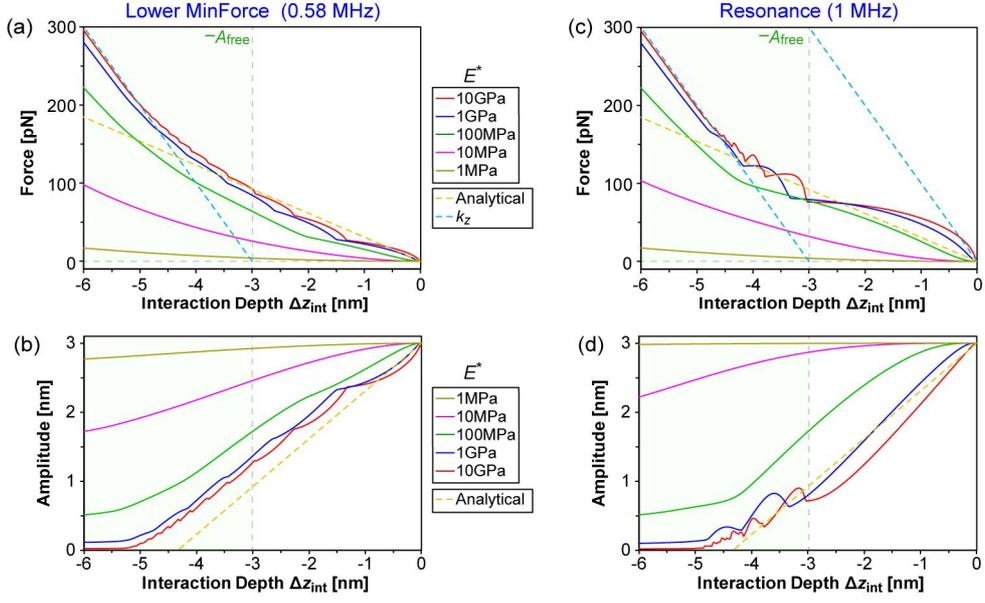

**FIG. 4. (a–d)** Interaction depth dependence of the average force (a,c) and amplitude (b,d) on sample with various Young's moduli when excited at the lower MinForce (a,b) and resonance (c,d) frequencies. The vertical broken lines indicate the position of $-A_{\text{free}}$ ($= -3.0$ nm$_{\text{p–0}}$). In Fig. 4(c), two lines with a gradient of $k_{\text{cl}}$ are plotted, starting from $\Delta z_{\text{int}} = 0$ and $\Delta z_{\text{int}} = -A_{\text{free}}$. The analytical calculations are represented by Eqs. (50) and (52) at the lower MinForce frequency.

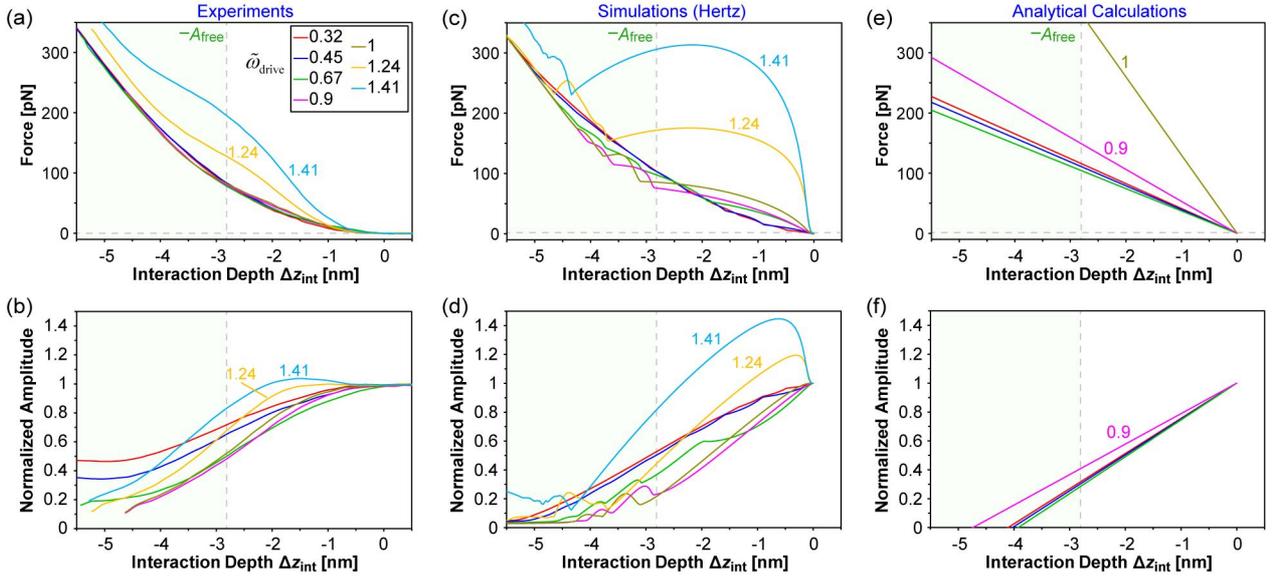

**FIG. 5.** (**a–f**) Comparison of experimental results (a,b), simulations (c,d), and analytical calculations (e,f) for the interaction depth dependence of the average force (a,c,e) and normalized amplitude (b,d,f). The analytical calculations are represented by Eqs. (49) and (51). The vertical broken lines indicate the position of $-A_{\text{free}}$ ($=-2.8$ nm$_{\text{p–0}}$).

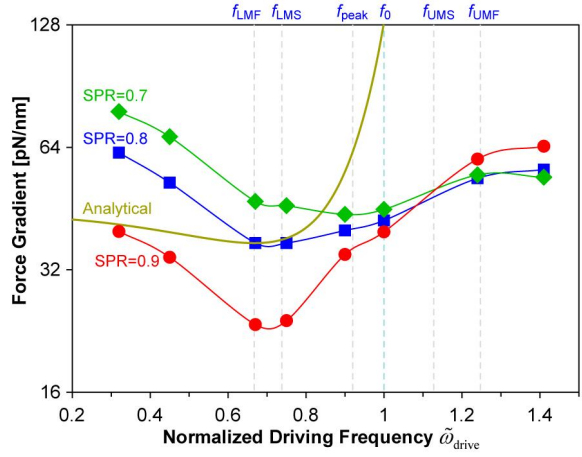

**FIG. 6.** Comparison of force gradients from analytical calculations and experiments with various setpoint ratios (SPRs), shown as a function of normalized driving frequency. The standard error of the experimental data, ranging from 0.37 to 0.66 pN/nm, is comparable to the line thickness; thus, error bars are omitted.

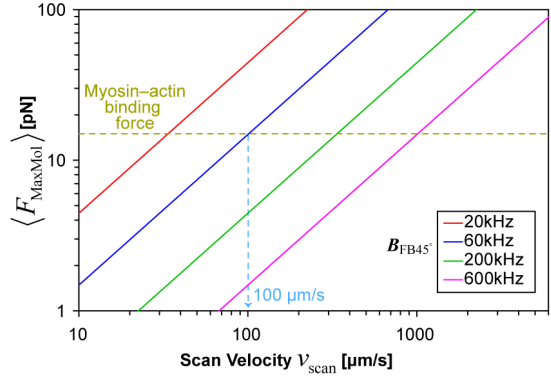

**FIG. 7.** Scan velocity dependence of maximum tip–sample interaction force ($\langle F_{\mathrm{MaxMol}} \rangle$) with various $B_{\mathrm{FB45°}}$ (Eq. (34)). The myosin–actin binding force (15 pN) is displayed as a reference. The scan velocity, determined from $B_{\mathrm{FB45°}}$ assumed in the previous HS-AFM experimental condition, is indicated by the broken arrow. $\langle F_{\mathrm{limit}} \rangle$ is calculated to be 157 pN, exceeding the Y-axis maximum and thus not shown.

**Acknowledgments**

This work was supported by PRESTO, Japan Science and Technology Agency (JST) [JPMJPR20E3 and JPMJPR23J2 to K.U.]; and KAKENHI, Japan Society for the Promotion of Science [21K04849 (to K.U.), 20H00327, and 24H00402 (to N.K.)].

**Author contributions**

K. U. constructed the theories, derived equations, and wrote the manuscript; and N. K. supervised the study.

**Data availability**

The data that support the findings of this study are available from the corresponding author upon reasonable request.